\documentclass[12pt]{article}
\usepackage{a4}
\usepackage{amsfonts,amssymb,amsmath}
\usepackage{graphicx}
\usepackage{cite}

\begin{document}

\title{Noncommutativity due to spin}
\author{M.Gomes\thanks{e-mail: mgomes@fma.if.usp.br}, V.G. Kupriyanov\thanks{e-mail:
vladislav.kupriyanov@gmail.com}, A.J. da Silva\thanks{e-mail: ajsilva@fma.if.usp.br} \\
%EndAName
Instituto de F\'{\i}sica, Universidade de S\~{a}o Paulo, Brazil\\
}
\maketitle

\begin{abstract}
Using the Berezin-Marinov pseudoclassical formulation of spin particle we
propose a classical model of spin noncommutativity. In the nonrelativistic
case, the Poisson brackets between the coordinates are proportional to the
spin angular momentum. The quantization of the model leads to the
noncommutativity with mixed spacial and spin degrees of freedom. A modified
Pauli equation, describing a spin half particle in an external e.m. field is
obtained. We show that nonlocality caused by the spin noncommutativity depends
on the spin of the particle; for spin zero, nonlocality does not appear, for
spin half, $\Delta x\Delta y\geq\theta^{2}/2$, etc. In the relativistic case
the noncommutative Dirac equation was derived. For that we introduce a new
star product. The advantage of our model is that in spite of the presence of
noncommutativity and nonlocality, it is Lorentz invariant. Also, in the
quasiclassical approximation it gives noncommutativity with a nilpotent parameter.

\end{abstract}

\section{Introduction}

The idea of using noncommutative (NC) coordinates in quantum mechanics
appeared a long time ago. In \cite{Peierls} noncommutative coordinates were
used to describe the charged particle in the strong magnetic field and in the
presence of the weak electric potencial. In the last decade, remotivated by
string theory arguments \cite{SW}, the subject gained a lot of interest and
has been studied extensively (see e.g. \cite{NCreviews} and \cite{NCQM} for
reviews on noncommutativity in QFT and QM, respectively). The canonical
noncommutative space can be realized by the coordinate operators $\hat{x}%
^{i},$ satisfying commutation relations $\left[  \hat{x}^{i},\hat{x}%
^{j}\right]  =i\theta^{ij},$ where $\theta^{ij\,}$ is an antisymmetric
constant matrix.

Recently, other types of noncommutativity, different from the canonical, have
also been considered. Thus, in \cite{GK09} it was proposed a model of position
dependent noncommutativity in quantum mechanics. In \cite{DN} a model of
dynamical noncommutativity was discussed. The authors of \cite{Gamboa1} have
proposed a three-dimensional noncommutative quantum mechanical system with
mixing spacial and spin degrees of freedom. The noncommutative spatial
coordinates $\hat{x}^{i}$, the conjugate momenta $\hat{p}_{i}$, and the spin
variables $\hat{s}^{i}$ were supposed to satisfy the non-standard Heisenberg
algebra:%
\begin{align}
\left[  \hat{x}^{i},\hat{x}^{j}\right]   &  =i\theta^{2}\varepsilon^{ijk}%
\hat{s}^{k},\label{1}\\
\left[  \hat{x}^{i},\hat{p}_{j}\right]   &  =i\delta_{j}^{i},\ \ \ \left[
\hat{p}_{i},\hat{p}_{j}\right]  =0,\nonumber\\
\left[  \hat{x}^{i},\hat{s}^{j}\right]   &  =i\theta\varepsilon^{ijk}\hat
{s}^{k},\ \ \ \ \left[  \hat{s}^{i},\hat{s}^{j}\right]  =i\varepsilon
^{ijk}\hat{s}^{k},\nonumber
\end{align}
where $\theta$ is the parameter of noncommutativity (a real number). We will
call it spin noncommutativity. Later, in \cite{Gamboa2} it was elaborated an
approach to the Bose-Einstein condensation theory, based on the spin noncommutativity. 
Note that in $2+1$ dimensions the relation between anyon spin and noncommutativity was discussed in \cite{JN}.

In the present work we will discuss some questions regarding the physical
meaning and mathematical formulation of the spin noncommutativity (\ref{1}).

It is known that canonical NCQM \cite{NCQM} can be obtained as a result of
quantization of classical models, see e.g., \cite{Lukierski2}%
-\cite{Deriglazov}. The corresponding action functional appears as an
effective action in path integral representation of NCQM \cite{Acatrinei}%
-\cite{Girotti} and can be used for study of global and local symmetries of
the system \cite{Kup8}, etc. The first question is if there exists a classical
model, which after quantization lead to the spin noncommutativity.

Another question is connected with nonlocality. Usually, noncommutativity
means the presence of nonlocality, i.e., nontrivial uncertainty relations
between the coordinates,%
\[
\Delta x^{i}\Delta x^{j}\geq\text{something}\neq0.
\]
The question is what is the form of nonlocality caused by spin
noncommutativity? Also, it is important to understand how the presence of the
spin noncommutativity can affect the relations between spin and statistics;
however, we will not discuss it in the present paper.

The last point, we would like to discuss here, is how to formulate a
consistent relativistic version of spin noncommutativity. In particular, we
will obtain the modification of the Dirac equation in the case of spin noncommutativity.

The paper is organized as follows. In the Sec. 2 we discuss the classical
model. The quantization of this model, constructed in Sec. 3, leads to the
modified Pauli equation and not to the Schrodinger one. We also discuss the
possibility of relativistic generalization of our model.

\section{Particle spin dynamics and its noncommutative deformation}

In \cite{BM} Berezin and Marinov have proposed\footnote{The similar model was considered independently in \cite{BCL}.} a classical model of the spin
1/2 particle, involving Grassmann degrees of freedom. In the nonrelativistic
case, the classical mechanics of a particle with spin is constructed in the
phase superspace, consisting of the six-dimmensional orbital subspace $\left(
x^{i},p_{i}\right)  ,\ i=1,2,3$, and three-dimensional spin Grassmann subspace
$\xi^{i}$, $\xi^{i}\xi^{j}+\xi^{j}\xi^{i}=0.$

The Poisson bracket between two arbitrary functions $f$ and $g$ of the
Grassmann variables is determined as follows,%
\begin{equation}
\left\{  f\left(  \xi\right)  ,g\left(  \xi\right)  \right\}  =-i\left(
f\overleftarrow{\partial_{k}}\right)  \left(  \overrightarrow{\partial_{k}%
}g\right)  . \label{3}%
\end{equation}
This Poisson bracket is antisymmetric if both functions are even elements of
the Grassmann algebra, and if one of them is an even element while other one
is an odd element. If both functions are odd elements, the Poisson bracket is
symmetric. For the canonical variables, the Poisson brackets are%
\begin{equation}
\left\{  \xi^{k},\xi^{l}\right\}  =-i\delta^{kl},\ \ \ \ \left\{  x^{k}%
,p_{l}\right\}  =\delta_{l}^{k}.
\end{equation}

The rotation group in the Grassmann subspace is generated by the spin angular
momentum%
\begin{align}
&  S^{i}=-\frac{i}{2}\varepsilon^{ijk}\xi^{j}\xi^{k},\label{5}\\
&  \left\{  S^{i},\xi^{j}\right\}  =\varepsilon^{ijk}\xi^{k},\ \ \ \ \left\{
S^{i},S^{j}\right\}  =\varepsilon^{ijk}S^{k}.\nonumber
\end{align}
The orbital angular momentum $L^{i}=\varepsilon^{ikl}x^{k}p^{l}$ generates the
rotation group in the orbital subspace,%
\[
\left\{  L^{i},x^{j}\right\}  =\varepsilon^{ijk}x^{k},\ \ \ \ \left\{
L^{i},L^{j}\right\}  =\varepsilon^{ijk}L^{k}.
\]
The complete angular momentum is determined as being%
\[
\mathbf{J}=\mathbf{L}+\mathbf{S},\ \ \left\{  J^{i},J^{j}\right\}
=\varepsilon^{ijk}J^{k}.\
\]

The classical Hamiltonian action of the model reads%
\begin{equation}
S_{0}=\int dt\left[  \mathbf{p\dot{x}}-\frac{i}{2}\mathbf{\xi\dot{\xi}%
}-H\left(  x,p,\xi\right)  \right]  , \label{6}%
\end{equation}
where%
\begin{equation}
H\left(  x,p,\xi\right)  =\frac{\mathbf{p}^{2}}{2}+V_{0}\left(  x\right)
+\left(  \mathbf{LS}\right)  V_{1}\left(  x\right)  +\mathbf{SB}\left(
x\right)  , \label{7}%
\end{equation}
$V_{0}\left(  x\right)  $ and $V_{1}\left(  x\right)  $ are potential
functions, and $\mathbf{B}\left(  x\right)  $ is a vector field. The term with
$V_{1}$ in (\ref{6}) is the spin-orbit interaction. The quantization of the
theory (\ref{6}) leads to the Pauli equation describing quantum
nonrelativistic spin $1/2$ particle.

Now, let us deform the above model to obtain nonzero Poisson brackets between
the coordinates, which may lead to noncommutativity after quantization. The
simplest way to do it is to\ mix coordinates and momenta \cite{Deriglazov},
$x_{NC}^{i}=x^{i}-1/2\theta^{ij\,}p_{j}$. However, this breaks symmetries of
the system, e.g., rotational symmetry, as $x_{NC}^{i}$ is not a vector anymore
(it does not transform as a vector, since $\theta^{ij\,}$ is a constant
matrix). To preserve rotational symmetry, one can mix coordinates and spin
angular momentum:%
\begin{equation}
\tilde{x}^{i}=x^{i}+\theta S^{i}. \label{8}%
\end{equation}
These new coordinates $\tilde{x}^{i}$, like the old ones are even and
transform like a vector,%
\begin{equation}
\left\{  J^{i},\tilde{x}^{j}\right\}  =\varepsilon^{ijk}\tilde{x}^{k}.
\label{9}%
\end{equation}
The nonvanishing Poisson brackets, involving new coordinates, are%
\begin{align}
\left\{  \tilde{x}^{i},\tilde{x}^{j}\right\}   &  =\theta^{2}\varepsilon
^{ijk}S^{k},\ \ \ \left\{  \tilde{x}^{i},p_{j}\right\}  =\delta_{j}%
^{i},\ \ \ \label{10}\\
\left\{  \tilde{x}^{i},\xi^{j}\right\}   &  =\theta\varepsilon^{ijk}\xi
^{k},\ \ \ \left\{  \xi^{k},\xi^{l}\right\}  =-i\delta^{kl}.\nonumber
\end{align}
Let us suppose that $\tilde{x}^{i}$ are "physical", i.e., observable
coordinates. We note that the center of mass coordinates in the Schrodinger Zitterbewegung 
problem satisfy similar commutation relations as $\tilde{x}^{i}$, see \cite{Horvathy}.  
One can then treat Poisson brackets (\ref{10}) as fundamental
Poisson brackets of a new theory in a phase superspace $\left(  \tilde
{x},p,\xi\right)  $. The graded version of the Jacobi identity in the deformed
theory can be easily verified. The Hamiltonian of the deformed theory is
$H\left(  \tilde{x},p,\xi\right)  $, where $H\left(  x,p,\xi\right)  $ was
determined in (\ref{7}).

In fact, this deformation is equivalent to the addition of new terms in the
action (\ref{6}), which disappear in the limit $\theta\rightarrow0$. However,
since we already have a consistent Hamiltonian formulation, which is necessary
for the quantization, the exact form of these additional terms is
immaterial.\footnote{The corresponding action functional can be constructed
along the lines described in \cite{GK07}, taking into account the presence of
the Grassmann variables.}

\section{Quantization}

In the course of quantization we replace the Poisson brackets (\ref{10})
between the canonical variables by the commutator (anticommutator) of the
corresponding operators%
\begin{align}
&  \left[  \hat{x}^{i},\hat{x}^{j}\right]  =i\theta^{2}\varepsilon^{ijk}%
\hat{s}^{k},\ \ \ \left[  \hat{x}^{i},\hat{p}_{j}\right]  =i\delta_{j}%
^{i},\label{11}\\
&  \left[  \hat{x}^{i},\hat{\xi}^{j}\right]  =i\theta\varepsilon^{ijk}\hat
{\xi}^{k},\ \ \ \left[  \hat{\xi}^{i},\hat{\xi}^{j}\right]  _{+}=\delta
^{ij}.\nonumber
\end{align}
Renormalizing the operators $\hat{\xi}^{i}=\hat{\sigma}^{i}/\sqrt{2},$one gets
the Clifford algebra with three generators%
\begin{equation}
\left[  \hat{\sigma}^{i},\hat{\sigma}^{j}\right]  _{+}=2\delta^{ij}.
\end{equation}
The only irreducible representation of this algebra is two-dimensional, it can
be realized by the Pauli matrices $\sigma^{i}$. Consequently,%
\begin{equation}
\hat{s}^{i}=-\frac{i}{2}\varepsilon^{ijk}\hat{\xi}^{j}\hat{\xi}^{k}=\frac
{1}{2}\sigma^{i}. \label{12}%
\end{equation}
One can see that the commutation relations involving the spatial coordinates
$\hat{x}^{i}$, the conjugate momenta $\hat{p}_{i}$, and the spin variables
$\hat{s}^{i}$ are exactly those in (\ref{1}), as postulated in \cite{Gamboa1}.
However, we have obtained these commutation relations as result of a
consistent quantization of a corresponding classical theory.

The representation of the quantum algebra (\ref{1}) is%
\begin{equation}
\hat{x}^{i}=x^{i}\mathbf{I}+\theta\hat{s}^{i},\ \ \ \hat{p}_{i}=-i\partial
_{i}\mathbf{I}, \label{13}%
\end{equation}
where $\mathbf{I}$ is the $2\times2$ unit matrix, and $\hat{s}^{i}$ are
determined in (\ref{12}). The modified Pauli equation, describing a
nonrelativistic spinning particle in an external electromagnetic field is%
\begin{equation}
i\partial_{t}\varphi=\hat{H}\left(  \hat{x},\hat{p},\hat{\xi}\right)  \varphi,
\label{14}%
\end{equation}
where $\varphi$ is a Pauli spinor and the Hamiltonian is given in (\ref{7}).

According to (\ref{1}) one has the uncertainty relations%
\begin{equation}
\Delta x^{i}\Delta x^{j}\geq\theta^{2}\varepsilon^{ijk}\left\vert \left\langle
\Psi\right\vert \hat{s}^{k}\left\vert \Psi\right\rangle \right\vert ,
\label{15}%
\end{equation}
where $\left\vert \Psi\right\rangle $ is a given state. Note, that since the
operators $\hat{s}^{k}$ do not commute, one can not measure simultaneously
eigenvalues for all operators $\hat{s}^{k}$, one has to choose one of them,
e.g., $\hat{s}_{z}$. If the particle has spin zero, then $\hat{s}%
^{k}\left\vert \Psi\right\rangle =0,$ there is no nonlocality in this case.
For the spin $s$ different from zero, one has:
\[
\hat{s}_{z}\left\vert \Psi\right\rangle =s_{z}\left\vert \Psi\right\rangle
,\ \ \ \ s_{z}=-s,-s+1,...,s.
\]
Substituting this in (\ref{15}) one has%
\[
\Delta x\Delta y\geq\theta^{2}\left\vert s_{z}\right\vert .
\]
For the particle with the spin $s=1/2,$%
\[
\Delta x\Delta y\geq\frac{\theta^{2}}{2}.
\]
So, for the spin noncommutativity, nonlocality is proportional to the quantum
number $s_{z}$, i.e., depends on the spin\ of the particle. Physically one can
interpret this result as follows: the maximal precision to localize the
particle depends on its spin.

\section{Relativistic generalization}

In the relativistic case, the Hamiltonian form of the Berezin-Marinov action
is%
\begin{align}
&  S=\int_{\tau_{i}}^{\tau_{f}}\left[  p_{\mu}\dot{x}^{\mu}-\frac{i}{2}%
\xi_{\mu}\dot{\xi}^{\mu}+\frac{i}{2}\xi^{5}\dot{\xi}^{5}-\frac{i}{2}\chi
T_{1}-\lambda T_{2}\right]  d\tau,\label{16}\\
&  T_{1}=\xi^{\mu}\left(  p_{\mu}+eA_{\mu}\right)  +m\xi^{5},\ \ T_{2}=\left(
p_{\mu}+eA_{\mu}\right)  ^{2}-m^{2}+ieF_{\mu\nu}\xi^{\mu}\xi^{\nu},\nonumber
\end{align}
here $\xi^{\mu},~\xi^{5}$ are Grassmann variables, describing spin degrees of
freedom, $\lambda$ and $\chi$ are Lagrange multipliers, $\lambda$-commuting
and $\chi$-anticommuting. Nonvanishing Poisson brackets between the canonical
variables are%
\begin{equation}
\left\{  x^{\mu},p^{\nu}\right\}  =g^{\mu\nu},\ \ \left\{  \xi^{\mu},\xi^{\nu
}\right\}  =-ig^{\mu\nu},\ \ \left\{  \xi^{5},\xi^{5}\right\}  =i,\label{17}%
\end{equation}
where $g^{\mu\nu}=$diag$\left(  1,-1,-1,-1\right)  $. Also, one has two
first-class constraints:%
\begin{equation}
T_{1}=0,\ \ T_{2}=0.\label{18}%
\end{equation}

Observe that $T_{1}$ is an odd element of the Grassmann algebra, therefore the
Poisson bracket of $T_{1}$ with $T_{1}$ is not zero, but
\begin{equation}
\left\{  T_{1},T_{1}\right\}  =-iT_{2}, \label{19}%
\end{equation}
By its definition, $T_{2}$ is even, so that $\left\{  T_{2},T_{2}\right\}
=0,$ and%
\begin{equation}
\left\{  T_{2},T_{1}\right\}  =i\left\{  \left\{  T_{1},T_{1}\right\}
,T_{1}\right\}  \equiv0, \label{20}%
\end{equation}
due to the Jacobi identity. Thus, we have proved that (\ref{18}) are indeed
first-class constraints.

Generators of the Lorentz group $J_{\mu\nu}$ are defined as%
\begin{align}
J^{\mu\nu}  &  =L^{\mu\nu}+S^{\mu\nu},\label{21}\\
L^{\mu\nu}  &  =x^{\mu}p^{\nu}-x^{\nu}p^{\mu},\ \ S^{\mu\nu}=-i\xi^{\mu}%
\xi^{\nu}.\nonumber
\end{align}
In the classical theory%
\begin{align}
\left\{  L^{\mu\nu},x^{\lambda}\right\}   &  =g^{\mu\lambda}x^{\nu}%
-g^{\nu\lambda}x^{\mu},\label{22}\\
\left\{  S^{\mu\nu},\xi^{\lambda}\right\}   &  =g^{\mu\lambda}\xi^{\nu}%
-g^{\nu\lambda}\xi^{\mu}.\nonumber
\end{align}

To construct relativistic generalization of the spin type noncommutativity we
introduce new coordinates%
\begin{equation}
\tilde{x}^{\mu}=x^{\mu}+\theta W^{\mu},\label{23}%
\end{equation}
where%
\[
W^{\mu}=\frac{1}{2}\varepsilon^{\mu\nu\rho\sigma}p_{\nu}J_{\rho\sigma}%
=\frac{1}{2}\varepsilon^{\mu\nu\rho\sigma}p_{\nu}S_{\rho\sigma}%
\]
is the Pauli-Lubanski vector. By the definition, $\tilde{x}^{\mu}$ is an even
element of the Grassmann algebra, and it transforms like a vector,%
\begin{equation}
\left\{  J^{\mu\nu},\tilde{x}^{\lambda}\right\}  =g^{\mu\lambda}\tilde{x}%
^{\nu}-g^{\nu\lambda}\tilde{x}^{\mu}.\label{24}%
\end{equation}
The Poisson brackets involving new coordinates are%
\begin{align}
\left\{  \tilde{x}^{\mu},\tilde{x}^{\nu}\right\}   &  =-\theta\varepsilon
^{\mu\nu\rho\sigma}S_{\rho\sigma}-\frac{\theta^{2}}{2}\varepsilon^{\mu\nu
\rho\sigma}W_{\rho}p_{\sigma},\ \ \ \label{25}\\
\left\{  \tilde{x}^{\mu},p^{\nu}\right\}   &  =g^{\mu\nu},\ \ \left\{
\xi^{\mu},\xi^{\nu}\right\}  =-ig^{\mu\nu},\ \ \left\{  \xi^{5},\xi
^{5}\right\}  =i,\ \nonumber\\
\left\{  \tilde{x}^{\mu},\xi^{\nu}\right\}   &  =-\theta\varepsilon^{\mu
\nu\rho\sigma}p_{\rho}\xi_{\sigma}.\ \ \nonumber
\end{align}
Again, we treat coordinates $\tilde{x}^{\mu}$ as "physical" coordinates and
Poisson brackets (\ref{25}) as the fundamental Poisson brackets of a new
theory in a phase superspace $\left(  \tilde{x},p,\xi\right)  $. The
constraints (\ref{18}) should be modified. We postulate the form of the first
constraint as%
\begin{equation}
\tilde{T}_{1}=\xi^{\mu}\left(  p_{\mu}+eA_{\mu}\left(  \tilde{x}\right)
\right)  +m\xi^{5}=0.\label{26}%
\end{equation}
As in undeformed case, it is an odd element of the Grassmann algebra, since
$\tilde{x}^{\mu}$ is even. Following (\ref{19}) we determine the second
constraint as
\begin{align}
&  \tilde{T}_{2}=i\left\{  \tilde{T}_{1},\tilde{T}_{1}\right\}  =0,\label{27}\\
&  \tilde{T}_{2}=\left(
p_{\mu}+eA_{\mu}\right)  ^{2}-m^{2}+ie\tilde{F}_{\mu\nu}\xi^{\mu}\xi^{\nu
}+2ie\left\{  \xi^{\mu},A_{\nu}\right\}  \left(  p_{\mu}+eA_{\mu}\right)
\xi^{\nu},\nonumber\\
&  \tilde{F}_{\mu\nu}=\frac{1}{e}\left\{  p_{\mu}+eA_{\mu}\left(  \tilde
{x}\right)  ,p_{\nu}+eA_{\nu}\left(  \tilde{x}\right)  \right\}  .\nonumber
\end{align}
It is even, since the Poisson bracket of two odd elements is always even.
Therefore, $\left\{  \tilde{T}_{2},\tilde{T}_{2}\right\}  =0$, and%
\begin{equation}
\left\{  \tilde{T}_{2},\tilde{T}_{1}\right\}  =i\left\{  \left\{  \tilde
{T}_{1},\tilde{T}_{1}\right\}  ,\tilde{T}_{1}\right\}  \equiv0,
\end{equation}
due to the Jacobi identity. Thus, the modified constraints $\tilde{T}_{1}=0$
and $\tilde{T}_{2}=0$ are again first-class constraints.

\section{Noncommutative Dirac equation}

After quantization the Poisson brackets (\ref{25}) will fix the commutation
(anticommutation) relations between the corresponding operators
\begin{align}
&  \left[  \hat{x}^{\mu},\hat{x}^{\nu}\right]  =-i\theta\varepsilon^{\mu
\nu\rho\sigma}\hat{S}_{\rho\sigma}+\frac{i\theta^{2}}{2}\varepsilon^{\mu
\nu\rho\sigma}\hat{W}_{\rho}\hat{p}_{\sigma},\label{28}\\
&  \left[  \hat{x}^{\mu},\hat{p}^{\nu}\right]  =ig^{\mu\nu},\ \ \left[
\hat{\xi}^{\mu},\hat{\xi}^{\nu}\right]  _{+}=g^{\mu\nu},\ \ \left[  \hat{\xi
}^{5},\hat{\xi}^{5}\right]  _{+}=-1,\nonumber\\
&  \left[  \hat{x}^{\mu},\hat{\xi}^{\nu}\right]  =-i\theta\varepsilon^{\mu
\nu\rho\sigma}\hat{\xi}_{\rho}\hat{p}_{\sigma}.\nonumber
\end{align}
The operators $\hat{\xi}^{\mu},\ \hat{\xi}^{5}$ are generators of the Clifford
algebra $C_{5}$. Its representation is four dimensional and is given by the
Dirac matrices:%
\begin{equation}
\hat{\xi}^{\mu}=i\gamma^{5}\gamma^{\mu}/\sqrt{2},\ \ \hat{\xi}^{5}=i\gamma
^{5}/\sqrt{2}. \label{gamma}%
\end{equation}
The representation of the operators of noncommutative coordinates $\hat
{x}^{\mu}$ and momenta $\hat{p}^{\mu}$ is
\begin{equation}
\hat{x}^{\mu}=x^{\mu}\mathbf{I}-\frac{i\theta}{2}\varepsilon^{\mu\nu
\alpha\beta}\hat{S}_{\alpha\beta}\partial_{\nu},\ \ \ \hat{p}_{\mu}%
=-i\partial_{\mu}\mathbf{I}, \label{29}%
\end{equation}
where $\mathbf{I}$ is $4\times4$ unit matrix, and%
\begin{equation}
\hat{S}_{\alpha\beta}=-\frac{i}{2}\left(  \hat{\xi}_{\alpha}\hat{\xi}_{\beta
}-\hat{\xi}_{\beta}\hat{\xi}_{\alpha}\right)  =-\frac{1}{4}\left(
\gamma_{\alpha}\gamma_{\beta}-\gamma_{\beta}\gamma_{\alpha}\right)  =\frac
{i}{2}\sigma_{\alpha\beta}. \label{30}%
\end{equation}
The first equation of (\ref{29}) is the analog of the Bopp shift; it can be
also represented as%
\begin{equation}
\hat{x}^{\mu}=x^{\mu}\mathbf{I}-\frac{i\theta}{2}\gamma^{5}\sigma^{\mu\nu
}\partial_{\nu}. \label{31}%
\end{equation}

Following \cite{KV} we define the star product through the Weyl symmetrically
ordered operator product as
\begin{equation}
\mathcal{W}(f\star g)=\mathcal{W}(f)\cdot\mathcal{W}(g), \label{starpr}%
\end{equation}
where%
\begin{equation}
\mathcal{W}(f)=\hat{f}\left(  \hat{x}\right)  =\int\frac{d^{4}k}{\left(
2\pi\right)  ^{4}}\tilde{f}\left(  k\right)  e^{-ik_{\mu}\hat{x}^{\mu}},
\label{2}%
\end{equation}
and $\tilde{f}(p)$ is the Fourier transform of $f$. This product is
associative due to the associativity of the operator products. Since,%
\[
\left[  -ik_{\alpha}x^{\alpha},k_{\mu}\theta\gamma^{5}\sigma^{\mu\nu}%
\partial_{\nu}/2\right]  =0,
\]
the exponential in the integral (\ref{2}) can be represented as%
\begin{equation}
e^{-ik_{\mu}\hat{x}^{\mu}}=e^{-ik_{\mu}x^{\mu}}e^{k_{\mu}\theta\gamma
^{5}\sigma^{\mu\nu}\partial_{\nu}/2}. \label{2a}%
\end{equation}
So,
\begin{equation}
\mathcal{W}(f)\cdot1=f\left(  x\right)  , \label{unity}%
\end{equation}
the result of the action of the polydifferential operator on a constant is a
function. The equations (\ref{starpr}) and (\ref{unity}) yield the following
formula
\begin{equation}
(f\star g)(x)=\mathcal{W}(f)g(x)=\hat{f}\left(  \hat{x}\right)  g(x)~,
\label{d3}%
\end{equation}
where the right hand side means an action of a polydifferential operator on a
function. The equation (\ref{d3}) can be written as%
\begin{align*}
&  \int\frac{d^{4}k}{\left(  2\pi\right)  ^{4}}\tilde{f}\left(  k\right)
e^{-ik_{\mu}x^{\mu}}e^{k_{\mu}\theta\gamma^{5}\sigma^{\mu\nu}\partial_{\nu}%
/2}g(x)\\
&  =fg+\sum_{n=1}^{\infty}\frac{\theta^{n}}{2^{n}n!}\int\frac{d^{4}k}{\left(
2\pi\right)  ^{4}}\tilde{f}\left(  k\right)  e^{-ik_{\mu}x^{\mu}}\left(
-ik_{\mu_{1}}\right)  ...\left(  -ik_{\mu_{n}}\right)  \gamma^{5}\sigma
^{\mu_{1}\nu_{1}}\partial_{\nu_{1}}...\gamma^{5}\sigma^{\mu_{n}\nu_{n}%
}\partial_{\nu_{n}}g(x).
\end{align*}
Finally we obtain the expression for the star product as%
\begin{equation}
f\star g=f\exp\left\{  \frac{i\theta}{2}\overleftarrow{\partial_{\mu}}%
\gamma^{5}\sigma^{\mu\nu}\overrightarrow{\partial_{\nu}}\right\}  g.
\label{def}%
\end{equation}

The first-class constraints (\ref{26}), (\ref{27}) are converted into
conditions on the physical states%
\begin{equation}
\hat{T}_{1}\psi=0,\ \ \ \hat{T}_{2}\psi=0,\label{32}%
\end{equation}
where some ordering should be specified. We choose the Weyl ordering. Using
the representation (\ref{gamma})-(\ref{31}) of the algebra (\ref{28}) one
writes the first equation of (\ref{32}) as%
\begin{equation}
\left[  i\gamma^{\mu}\left(  \partial_{\mu}+ieA_{\mu}\left(  x^{\mu}%
\mathbf{I}-\frac{i\theta}{2}\gamma^{5}\sigma^{\mu\nu}\partial_{\nu}\right)
\right)  -m\right]  \psi=0.\label{33}%
\end{equation}
Taking into account the definition of the star product (\ref{def}), the above
equation can be represented in the form%
\begin{equation}
\left[  i\gamma^{\mu}\left(  \partial_{\mu}+ieA_{\mu}\left(  x\right)
\right)  -m\right]  \star\psi=0.\label{34}%
\end{equation}
We call this equation as noncommutative Dirac equation. In contrast to the
case of canonical noncommutativity, it is a relativistic equation (in the
sense of special relativity), covariant under the Lorentz transformation%
\begin{equation}
x^{\mu}\rightarrow x^{\prime\mu}=\Lambda_{\nu}^{\mu}x^{\nu},\ \ \psi
\rightarrow\psi^{\prime}\left(  x^{\prime}\right)  =S\left(  \Lambda\right)
\psi\left(  x\right)  ,\ \ A^{\mu}\rightarrow A^{\prime\mu}\left(  x^{\prime
}\right)  =\Lambda_{\nu}^{\mu}A^{\nu}\left(  x\right)  ,
\end{equation}
where $S\left(  \Lambda\right)  $ belongs to the usual spinor representation
of the Lorentz group. This assertion follows by a direct use of the identities%
\begin{equation}
S^{-1}\gamma^{\mu}S=\Lambda_{\alpha}^{\mu}\gamma^{\alpha},\ \ S^{-1}%
\sigma^{\mu\nu}S=\Lambda_{\alpha}^{\mu}\Lambda_{\beta}^{\nu}\sigma
^{\alpha\beta}.
\end{equation}
The second equation of (\ref{32}) is a consequence of the first one, since
$\hat{T}_{2}=\left(  \hat{T}_{1}\right)  ^{2}$.

Note, that a quasiclassical approximation in the spin degrees of freedom
(e.g., a partial quantization of bosonic coordinates only) leads to the
noncommutativity with bifermionic NC parameter \cite{GV}:%
\begin{equation}
x^{\mu}\star x^{\nu}-x^{\nu}\star x^{\mu}=i\Theta^{\mu\nu},\ \ \Theta^{\mu\nu
}=i\ \theta\varepsilon^{\mu\nu\rho\sigma}\xi^{\rho}\xi^{\sigma}/2.
\end{equation}
Similar constructions also appeared in the context of nonanticommutative
superspace \cite{Seiberg}. Nilpotent noncommutativity can improve the
renormalizability properties of noncommutative theories, \cite{FGV}.

\section{Conclusions}

In the present paper we have derived a model of noncommutativity with mixed
spatial and spin degrees of freedom. For that we have constructed a consistent
deformation of the Berezin-Marinov pseudoclassical formulation of spin
particle. In the nonrelativistic case the deformed coordinates are the sum of
the initially commutative coordinates and the spin angular momentum,
$\tilde{x}^{i}=x^{i}+\theta S^{i}$. The Poisson brackets between the deformed
coordinates are proportional to the spin angular momentum, which leads to the
spin noncommutativity after quantization. In the relativistic case the
deformed coordinates are the sum of the commutative coordinates and the
Pauli-Lubanski vector, $\tilde{x}^{\mu}=x^{\mu}+\theta W^{\mu}$.

Also we have obtained the modified Pauli equation (in the nonrelativistic
case) and the noncommutative Dirac equation (in the relativistic case),
describing spin half particle in an external electromagnetic field in the
presence of the spin noncommutativity. Nonlocality in our model depends on the
spin of the particle.

We stress that the noncommutative Dirac equation (\ref{34}) is covariant under
Lorentz transformations. Therefore, it can be used as a basis for the
construction of a consistent relativistic noncommutative field theory. The next step in this way is to introduce the trace
functional on the algebra of the star product (\ref{32}) and to construct corresponding
action functional for the noncommutative Dirac field. Also, stell in the context of quantum mechanics, it would be
interesting to study phenomenological effects caused by the spin
noncommutativity on the examples of exact solvable QM models like Hydrogen
atom, Landau problem, Aharonov-Bohm effect, etc.

\section*{Acknowledgements}

V.G.K. acknowledges FAPESP for support. M.G. and A.J.S. thank FAPESP and CNPq for partial support.


\begin{thebibliography}{99}                                                                                               %


\bibitem {Peierls}R. Peierls, Z.Phys. \textbf{80} (1933) 763.

\bibitem {SW}N. Seiberg and E. Witten, JHEP 9909 (1999) 032.

\bibitem {NCreviews}M. Douglas, N. Nekrasov, Rev.Mod.Phys.\textbf{73} (2001)
977-1029; R. Szabo, Phys.Rept.\textbf{378} (2003) 207-299.

\bibitem {NCQM}C. Duval, P.A. Horvathy, Phys.Lett.\textbf{B479}(2000)284;
M.~Chaichian, M.M.~Sheikh-Jabbari and A.~Tureanu, Phys.Rev.Lett.\ \textbf{86}(2001)2716; J.~Gamboa, M.~Loewe and J.C.~Rojas, Phys.Rev.D\textbf{64 }(2001)067901; V.P. Nair, A.P. Polychronakos, Phys.Lett.\textbf{B505}(2001)267;
S. Bellucci, A. Nersessian, C. Sochichiu, Phys.Lett.\textbf{B522}(2001)345;
P.A. Horvathy, M.S. Plyushchay, JHEP06(2002)033;
A.F. Ferrari, M. Gomes, C.A. Stechhahn, Phys.Rev.\textbf{D76}(2007)085008;
P.G. Castro, B. Chakraborty, F. Toppan,\ J.Math.Phys.\textbf{49}(2008)082106.

\bibitem {GK09}M. Gomes, V.G. Kupriyanov, Phys.Rev.\textbf{D79} (2009) 125011.

\bibitem {DN}M. Gomes, V.G. Kupriyanov, A.J. da Silva, Dynamical
noncommutativity, arXiv:0908.2963 [hep-th].

\bibitem {Gamboa1}H. Falomir, J. Gamboa, J. Lopez-Sarrion, F. Mendez, P.A.G.
Pisani, Phys.Lett.\textbf{B680} (2009) 384-386.

\bibitem {Gamboa2}J. Gamboa, F. Mendez, Bose-Einstein condensation theory for
any integer spin: approach based in noncommutative quantum mechanics,
arXiv:0912.2645 [hep-th].

\bibitem {JN}R. Jackiw, V.P. Nair, Phys.Lett.\textbf{B480}(2000)237.

\bibitem {Lukierski2}J. Lukierski, P.C. Stichel., W.J. Zakrzewski, Annals
Phys.260 (1997) 224-249.

\bibitem {Duval}C. Duval, P. Horvathy, J. Phys. A \textbf{34} (2001) 10097.

\bibitem {Deriglazov}A.A. Deriglazov, Phys.Lett. B\textbf{555 }(2003) 83; JHEP
\textbf{0303} (2003) 021.

\bibitem {Acatrinei}C. Acatrinei, JHEP 09(2001)007.

\bibitem {Kup5}D.M.~Gitman, V.G.~Kupriyanov, Eur.Phys.J.\ C \textbf{54} (2008) 325.

\bibitem {Girotti}F.S. Bemfica, H.O. Girotti, Phys.Rev.\textbf{D77 }(2008)
027704, Phys.Rev.\textbf{D79} (2009) 125024;

\bibitem {Kup8}D.M.~Gitman, V.G.~Kupriyanov, J.Math.Phys. \textbf{51} (2010)
022905.

\bibitem {BM}F.~A. Berezin and M.~S. Marinov, Ann. Phys. \textbf{104 }(1977) 336.

\bibitem {BCL}A. Barducci, R. Casalbuoni, L. Lusanna, Nuovo Cimento \textbf{35A} (1976) 377.

\bibitem {Horvathy}P.A. Horvathy, Acta Phys.Pol. \textbf{B34} (2003) 2611.

\bibitem {GK07}D.M. Gitman and V.G. Kupriyanov, Eur.Phys.J.C \textbf{50}
(2007) 691-700.

\bibitem {KV}V.G. Kupriyanov, D.V. Vassilevich, Eur.Phys.J.C. \textbf{58}
(2008) 627-637.

\bibitem {GV}D.M. Gitman, D.V. Vassilevich, Mod.Phys.Lett.\textbf{A23} (2008) 887-893.

\bibitem {Seiberg}N. Seiberg, JHEP \textbf{0306} (2003) 010.

\bibitem {FGV}R. Fresneda, D.M. Gitman, D.V. Vassilevich,
Phys.Rev.\textbf{D78} (2008) 025004.

\end{thebibliography}
\end{document}